# Photoionization of Molecular Endohedrals

M. Ya. Amusia[1, 2][1], L. V. Chernysheva[2] and S. K. Semenov

[1]*Racah Institute of Physics, the Hebrew University, Jerusalem 91904, Israel*
[2]*A. F. Ioffe Physical-Technical Institute, St. Petersburg 194021, Russian Federation*
[3]*Saint Petersburg State University of Aerospace Instrumentation*, *St. Petersburg 190000, Russian Federation*

**Abstract:**
We calculate the photoionization cross-section of a molecular endohedral that we denote as $M@C_N$. We limit ourselves to two-atomic molecules. The consideration is much more complex than for atomic endohedrals because the system even for almost spherical $C_N$ has only cylindrical instead of spherical symmetry. On the other hand, $M@C_N$ is more interesting since the interelectron interaction in molecules is relatively stronger than in similar atoms.

We present here results of calculations of molecular hydrogen $H_2$ stuffed inside almost spherical fullerene $C_{60}$ – $H_2@C_{60}$. For comparison, we perform calculations also for atomic endohedral $He@C_{60}$. The results are obtained both in the single-electron Hartree-Fock approximation and with account of multi-electron correlations in the frame of so-called random phase approximation with exchange – RPAE.

The presence of the fullerenes shell results in prominent oscillations in the endohedrals photoionization cross section. The role of interelectron correlations becomes clear by comparing HF and RPAE results for $H_2@C_{60}$ and $He@C_{60}$ on the one side with that for $H_2$ and He, on the other.

**1**. We investigate here the photoionization of a molecular endohedral $M@C_N$ that consist of a molecule M stuffed inside a fullerene $C_N$ constructed of N carbon C atoms. There exist numerous investigations of photoionization of atomic endohedrals $A@C_N$, mainly theoretical, but also several experimental (see [1-4] and references therein). The theoretical studies include calculations not only of absolute cross section but also of photoelectron angular distribution and spin polarization. The calculations were performed in the frame of a variety of different approximations starting from very crude, up to quite complex and potentially accurate. However, as far as we know, no papers exist on $M@C_N$ photoionization, whereas the existence of such an endohedral is demonstrated.

The system $M@C_N$ is in principle much more complex than an atomic endohedral $A@C_N$, but at the same time more interesting. The interest comes from considerably more complex electronic structure of a molecule than that of an atom, and presence of internal degrees of freedom such as vibration and rotation that does not exist in atoms. Molecular photoionization is of special interest, since the photoelectron can come to the point of observation using different ways and after emission by different atoms of the molecule. Being relatively easy deformable, molecules can modify their shape under the action of the $C_N$ shell. In molecules the relative role of interelectron interaction is bigger than even in heavy atoms since they are weaker bound and the total positive charge of a molecule is distributed among several nuclei.

---

[1]amusia@vms.huji.ac.il

However, the non-spherical molecular shape that is a result of multi-center location of nuclei makes the calculations of molecular photoionization much more difficult than the calculation of atomic photoionization. For isolated atoms, as one-electron approach the Hartree-Fock approximation (HF) is usually employed. The interelectron correlations there are taken into account in the frame of the random phase approximation with exchange (RPAE). The loss of spherical symmetry already for simple two-atomic molecules results in the fact that even HF, not mentioning RPAE, becomes much more complex. This is why it became possible to apply HF and then RPAE to photoionization studies of two-atomic molecules only, neglecting molecular vibration and rotation [5, 6]. This limitations preserve axial and, in case of identical atoms, the reflection symmetry. To make calculations possible, it is necessary to neglect the vibrational and rotational degrees of freedom.

All these restrictions we apply in this Letter, considering as an object of photoionization a diatomic molecule, and as a concrete example the simplest of it: the hydrogen molecule $H_2$ stuffed inside an almost ideally spherical fullerene $C_{60}$, thus forming the molecular endohedral $H_2@C_{60}$. As the reference for comparison we choose an atomic endohedral $He@C_{60}$. The results for photoionization cross section of He we take from [7] and for $H_2$ from [5, 8]. We carry out all calculations in both HF and RPAE approximation. We assume that the spin-orbit interaction is inessential, so the spin and coordinate variables can be separated easily. The endohedral $C_{60}$ is represented by a spherically-symmetric square well potential $W(r)$ and a factor $G(\omega)$ that depends upon photon energy[2] $\omega$ and takes into account the dipole polarization of the fullerene $C_{60}$ by the incoming photon beam [8, 7].

**2.** As it was demonstrated in [9, 5], the most suitable for considering molecules that consist of two atoms are spheroidal coordinates $(\xi, \eta, \varphi)$ that replace the ordinary linear x, y and z and spherical $R, \theta, \varphi$ coordinates:

$$x = R(\xi^2 - 1)^{1/2}(1-\eta^2)^{1/2}\cos\varphi, \quad y = R(\xi^2 - 1)^{1/2}(1-\eta^2)^{1/2}\sin\varphi, \quad z = R\xi\eta, \tag{1}$$

with $R = D/2$ and $D$ being the interatomic distance in the molecule.

The general shape of HF and RPAE equations is the same as for atoms; however the atomic one-electron wave functions have to be substituted by the molecular wave functions. The main calculation difficulty originates from the fact that the angular momentum is no more a good quantum number. It is convenient to present the HF one-electron wave functions $\varphi_i(\mathbf{r})$ of the state $i$ that has a given projection of the angular momentum $m_i$ on the molecular axis as an expansion

$$\varphi_i(\mathbf{r}) = \sum_{\lambda=|m_i|}^{L} P_{i\lambda}(\xi) Y_{\lambda m_i}(\eta, \varphi), \tag{2}$$

---

[2] We employ here the atomic system of units $m = e = \hbar = 1$. Here $m$ is the electron mass; $e$ is its charge and $\hbar$ is the Planck constant.

Here $P_{i\lambda}(\xi)$ is the analog of usual radial function and $Y_{\lambda m_i}(\eta,\varphi)$ is the spherical harmonic. The maximal angular momentum $L$ of the expansion is determined in course of calculations.

The HF equations for the two-atomic molecule are the following [9, 8]:

$$\left[\frac{d}{d\xi}(\xi^2-1)\frac{d}{d\xi}-\frac{m_i^2}{\xi^2-1}-\lambda(\lambda+1)+2R(Z_1+Z_2)\xi+W(\xi^2+\eta^2)+\varepsilon_i\frac{R^2}{2}(\xi^2-d_{\lambda\lambda}^i)\right]P_{i\lambda}(\xi)$$

$$-\frac{R^2}{2}\sum_k P_{ik}(\xi)\sum_{j=1}^N (2-\delta_{ji})\sum_{\lambda'} Z_{jj,\lambda'}(\xi)\left[\xi^2 A_{0,m_i,m_i}^{\lambda',k,\lambda} - B_{0,m_i,m_i}^{\lambda',k,\lambda}\right]$$

$$+\frac{R^2}{2}\sum_{\substack{j=1\\(j\ne i)}}^N \sum_{k'} P_{jk'}(\xi)\sum_{\lambda'} Z_{ij,\lambda'}(\xi)\left[\xi^2 A_{m_{ij},m_j,m_i}^{\lambda',k',\lambda} - B_{m_{ij},m_j,m_i}^{\lambda',k',\lambda}\right] \quad (3)$$

$$=\varepsilon_i \frac{R^2}{2}\sum_{k\ne\lambda} d_{\lambda k}^i P_{ik}(\xi) - 2R(Z_2-Z_1)\sum_k c_{\lambda k}^i P_{ik}(\xi); \quad m_{ij}\equiv m_i-m_j, \quad \lambda=|m_i|,...,L.$$

Here

$$c_{\lambda\lambda'}^i = \iint Y_{\lambda m_i}(\eta,\phi)\eta Y_{\lambda' m_i}^*(\eta,\phi)d\eta d\phi, \quad d_{\lambda\lambda'}^i = \iint Y_{\lambda m_i}(\eta,\phi)\eta^2 Y_{\lambda' m_i}^*(\eta,\phi)d\eta d\phi,$$

$$A_{m,m',m''}^{l,l',l''} = \iint Y_{l,m}Y_{l',m'}Y_{l'',m''}^* d\eta d\varphi, \quad B_{m,m',m''}^{l,l',l''} = \iint Y_{l,m}Y_{l',m'}Y_{l'',m''}^* \eta^2 d\eta d\varphi,$$

Here $Z_1$ and $Z_2$ are the nuclear charges of atoms 1 and 2 that constitute a diatomic molecule.

Functions $Z_{ij,\lambda'}(\xi)$ are determined by the following equation [8]:

$$\left[\frac{d}{d\xi}(\xi^2-1)\frac{d}{d\xi}-\frac{m^2}{\xi^2-1}-\lambda(\lambda+1)\right]Z_{ij,\lambda}(\xi)$$

$$=-\pi R^2(-1)^m \sum_{k,k'} P_{i,k}(\xi)P_{j,k'}(\xi)\left(\xi^2 A_{m_j,-m_i,m}^{k',k,\lambda} - B_{m_j,-m_i,m}^{k',k,\lambda}\right), \quad \lambda=1\div L. \quad (4)$$

Similar, but not identical equations (with orthogonaliation parameters) determine the excited states and continuous spectrum wave functions $P_{f\lambda}(\xi)$ [8].

The operators of photon-electron interaction in the length form are given by the following relations:

$$d_{\pm 1}^L = \frac{R_z}{2}\sqrt{\frac{4\pi}{3}}(\xi^2-1)^{1/2}Y_{1,\pm 1}(\eta,\varphi), \quad d_0^L = \frac{R_z}{2}\xi\eta. \quad (5)$$

In velocity form a much more complex form determines the operator $d_\mu^V$.

The RPAE equation is similar to the atomic one and corresponds to a given projection of the dipole operator $\mu=\pm,0$:

$$\langle v|D_\mu(\omega)|i\rangle = \langle v|d_\mu|i\rangle + \sum_{i'<F}\sum_{v'>F}\left[\frac{\langle v'|D_\mu(\omega)|i'\rangle\langle i',v|U|v',i\rangle}{\omega-\varepsilon'_f+\varepsilon'_i+i\delta} - \frac{\langle i'|D_\mu(\omega)|v'\rangle\langle v',v|U|i',i\rangle}{\omega+\varepsilon'_f-\varepsilon'_i}\right], \quad (6)$$

Here $\langle i',v|U|v',i\rangle \equiv \langle i',v||\mathbf{r}-\mathbf{r}'|^{-1}|v',i-i,v'\rangle$; $v$ denotes the set $v \equiv \lambda_f, m_f, \varepsilon_f$, while $i \equiv m_i, \varepsilon_i$.

The following relations determine the contribution to the absolute photoionization cross section of the molecule due to $i \to \varepsilon$ transition, in atomic units and Megabarns (Mb), respectively [7]:

$$\sigma_{vi}(\omega_{vi}) = \frac{4\pi^2\alpha a_0^2}{3}\omega_{vi}\sum_\mu|<v|D_\mu(\omega_{vi})|i>|^2 = 2.689\cdot\omega_{vi}\sum_\mu|<v|D_\mu(\omega_{vi})|i>|^2,$$
$$\omega_{vi} \equiv \varepsilon_v - \varepsilon_i \quad (7)$$

As it was demonstrated in [10,], assuming that the fullerenes radius $R_C$ is much bigger that the molecular length and the atomic radius, one can take into account $C_N$ polarization using the formula

$$D_{iv}^{A@C_N}(\omega) = G^C(\omega)<v|D_\mu(\omega_{vi})|i> \equiv \left[1-\alpha_d^C(\omega)/R_C^3\right]<v|D_\mu(\omega_{vi})|i>. \quad (8)$$

Here $\alpha_d^C(\omega)$ is the $C_N$ dipole dynamic polarizability.

This expression leads to the following relation for the photoionization cross section of the endohedral with account of the photon beam modification due to fullerene $C_{60}$ polarization:

$$\sigma_{vi}^G(\omega_{vi}) \equiv \left|G^C(\omega_{vi})\right|^2 \sigma_{vi}(\omega_{vi}). \quad (9)$$

**3.** As we mentioned above, the concrete example of a diatomic molecule here is the molecular hydrogen, $H_2$. As a fullerene, our choice is an almost spherical fullerene $C_{60}$, so it is possible to model it by a spherical potential. It is important also that $C_{60}$ is one of the best studied fullerenes. We have calculated photoionization cross-sections of $H_2@C_{60}$ and, for comparison, $He@C_{60}$ at first in the frame of one-electron Hartree-Fock (HF) approximation, and then add interelectron correlations in the frame of the random phase approximation with exchange (RPAE). In both cases we compare the results obtained with that for isolated molecule $H_2$ and atom He. The necessary details about HF and RPAE equation and their solutions one can find in [7]. The next and last in this Letter is the inclusion of the effect of the fullerene polarization by the incoming photon beam using expression (8). As fullerenes potential W(*r*) we choose a square well (see parameters in [7, 8]). The upper value of *L* in (2) was *L*=5.

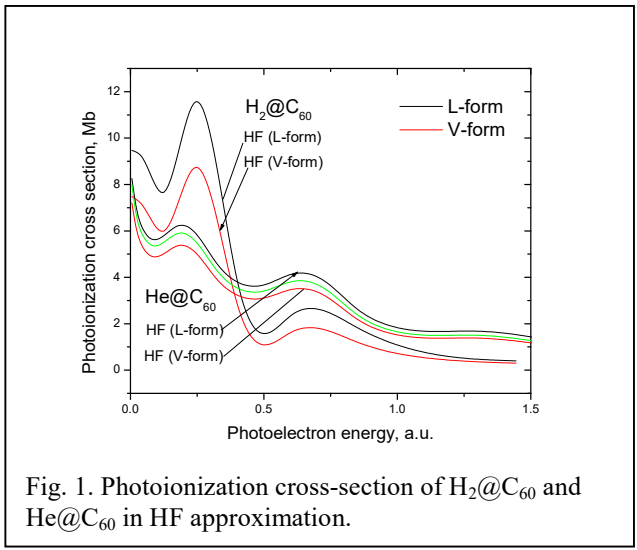

Fig. 1. Photoionization cross-section of $H_2@C_{60}$ and $He@C_{60}$ in HF approximation.

Fig. 1-4 present results of our calculations. Fig. 1 depicts the photoionization cross-section of $H_2@C_{60}$ and $He@C_{60}$ in HF approximation. As one could expect, the difference between results, calculated using "length" (5) and "velocity" forms of operator that describes photon-electron interaction, is small for He and not too big for $H_2$, however there it is essentially bigger than for He. The fullerenes shell adds oscillations in the cross-section that are much more prominent in $H_2$ case than for He.

Fig. 2 presents results for $H_2@C_{60}$ and $He@C_{60}$ in HF and adds RPAE data for $He@C_{60}$. We see impressive oscillations in $H_2@C_{60}$ and prominent suppression of the ionization cross section at threshold in $H_2@C_{60}$ as compared to that in isolated $H_2$ (see Fig. 3). The enhancement of oscillation in $H_2@C_{60}$ as compared to $He@C_{60}$ demonstrates the relative increase of the fullerene action upon the molecular endohedral photoionization cross section as compared to the atomic endohedral cross-section. This is a result of the bigger size of the molecule as compared to an atom with the same number of electrons and protons.

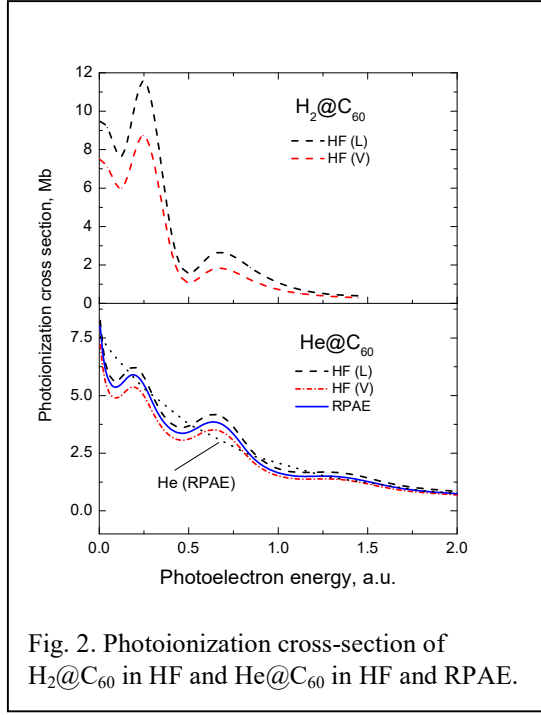

Fig. 2. Photoionization cross-section of $H_2@C_{60}$ in HF and $He@C_{60}$ in HF and RPAE.

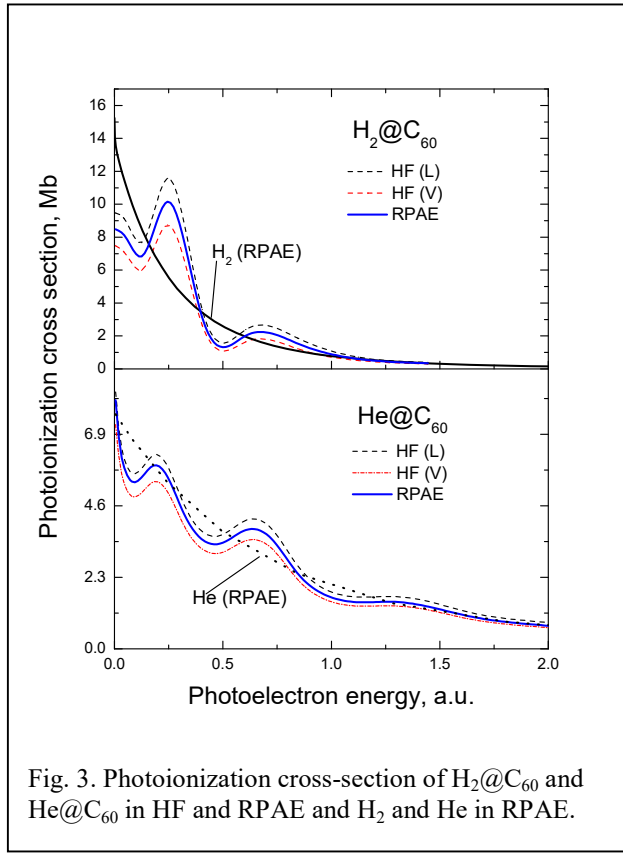

Fig. 3. Photoionization cross-section of $H_2@C_{60}$ and $He@C_{60}$ in HF and RPAE and $H_2$ and He in RPAE.

Fig. 3 present results for $H_2@C_{60}$ and $He@C_{60}$ in HF and RPAE as well as RPAE results for $H_2$ and He. One can see prominent differences between an isolated molecule and atom on the one side and respective endohedrals on the other. In RPAE, as it should be, the results obtained with "length" and "velocity" operators almost coincide. The oscillations in the cross section for molecular endohedral are much bigger than for the atomic endohedral. Note that for for isolated $H_2$ and He the dipole sum rule is valid

$$\sum_n f_n + (c/2\pi^2)\int_I^\infty \sigma(\omega)d\omega = N_e, \quad (10)$$

where $f_n$ are the oscillator strengths, $N_e$ is the number of electrons and *I* is the ionization potential of the ionizing object. Thus, the bigger value of $\sigma(I)$ for H$_2$ has to be compensated by faster decrease of $\sigma(\omega)$ with $\omega$ growth than in He. In endohedrals the sum rule could be violated due to essential redistribution in (10) between the fullerenes shell and stuffed object. This redistribution considerably affects the s-subshells of inner objects, usually diminishing their contribution [11].

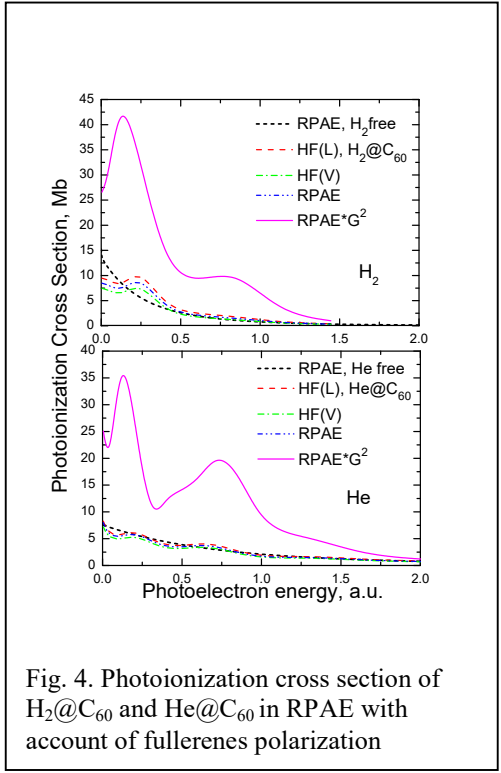

Fig. 4. Photoionization cross section of H$_2$@C$_{60}$ and He@C$_{60}$ in RPAE with account of fullerenes polarization

Fig. 4 presents the cross sections with account of $G^C(\omega)$ factor (9). We take the C$_{60}$ dipole polarizability from the paper [10]. It has two prominent maxima. As one can see, they dominate in $\sigma^G_{1s\varepsilon}(\omega_{1s\varepsilon})$. The endohedral photoionization cross section increases impressively. The H$_2$@C$_{60}$ cross-section is somewhat bigger than the cross-section of He@C$_{60}$, however due to specifics of the energy dependence of the G$^C$ factor, the difference between cross-sections for H$_2$@C$_{60}$ and He@C$_{60}$ decreases.

**4.** This research opens a new direction in photoionization studies of endohedrals. We present here formulas to calculate photoionization cross sections of molecular endohedrals. Concrete results are obtained for H$_2$@C$_{60}$ and He@C$_{60}$ as well as for H$_2$ and He in the frames of HF and RPAE approaches. All necessary computing codes we took from the system ATOM-M [8]. We found prominent oscillations in the photoionization cross section due to presence of the fullerene C$_{60}$ electron shells in the endohedrals. These oscillations in H$_2$@C$_{60}$ are somewhat stronger than in He@C$_{60}$. Most important, we demonstrate that molecular endohedral photoionization can be treated with the same good accuracy that it is achieved for atomic endohedrals.

Natural next steps would be investigation of angular and spin orientation distributions of photoelectrons and consideration of more complex diatomic molecules including those excited to vibrational levels, and those consisting of different atoms. In this area accurate experimentation is, perhaps, more difficult than performing calculations. But it is the experimental data that can determine convincingly that the developed theoretical approach is adequate to the considered problem.

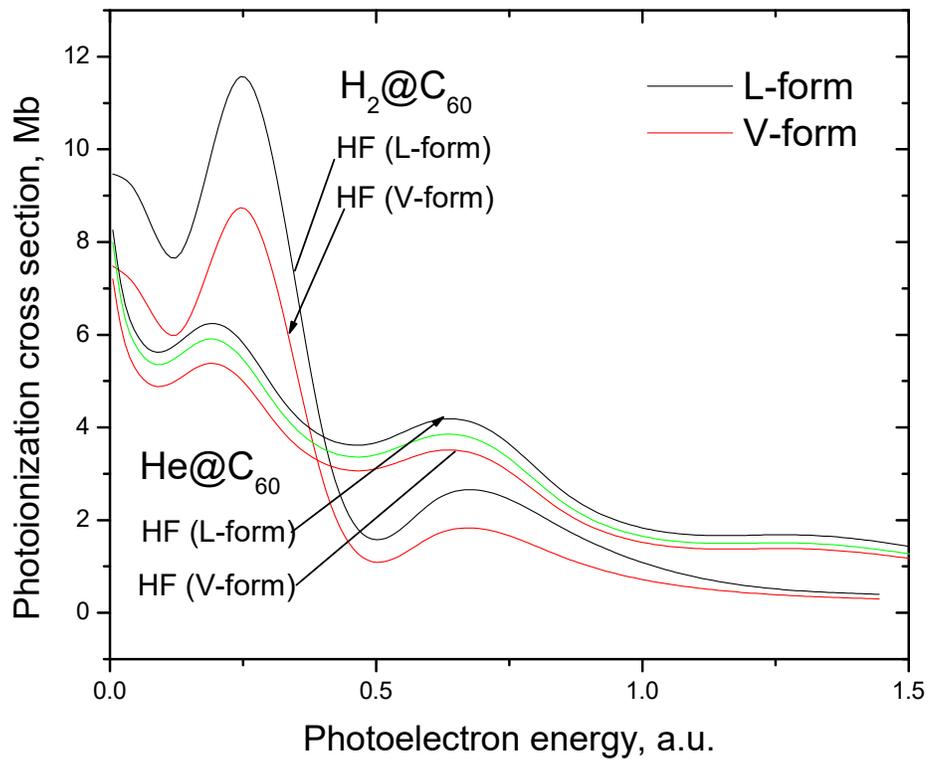

Fig. 1. Photoionization cross-section of $H_2@C_{60}$ and $He@C_{60}$ in HF approximation.

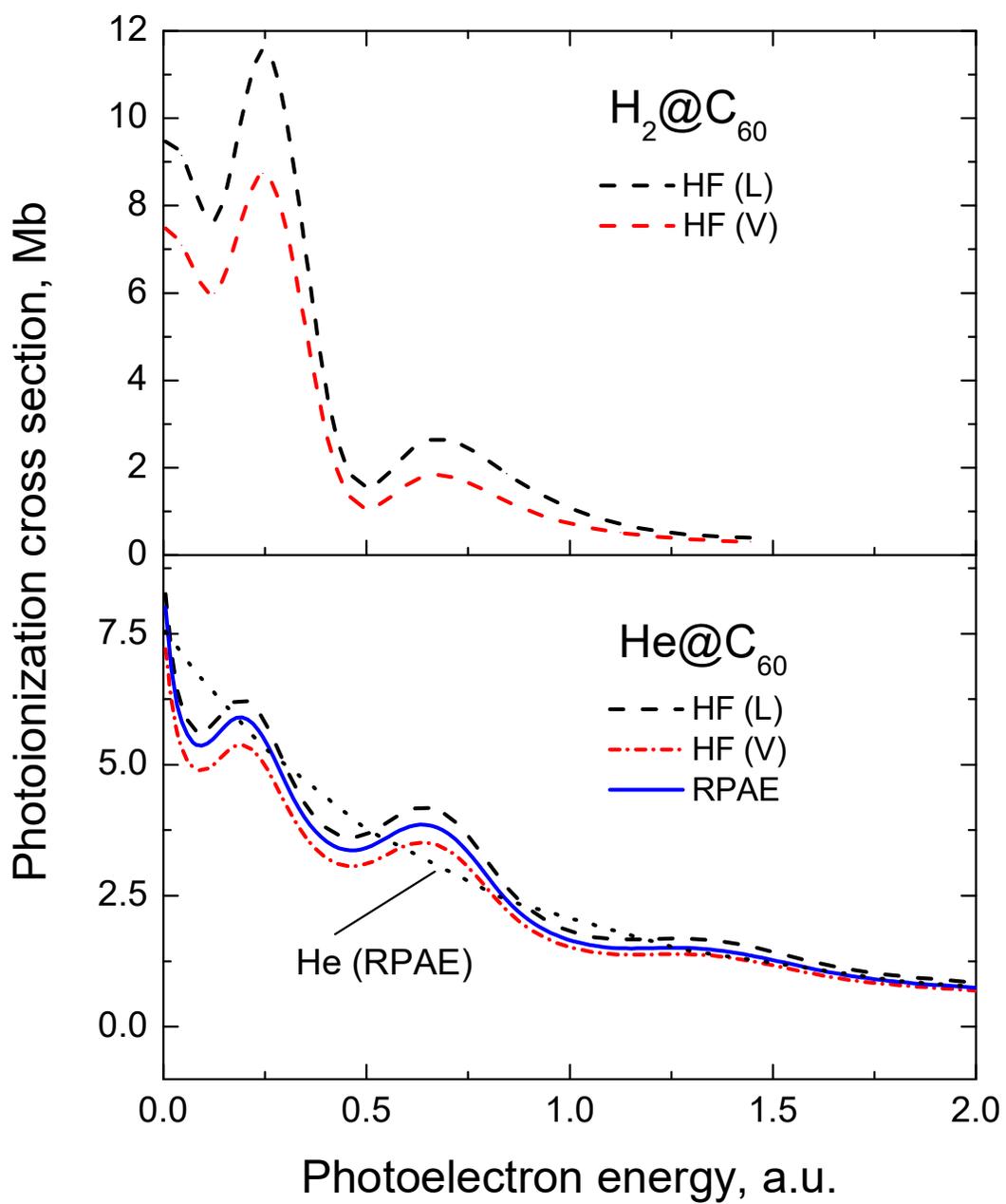

Fig. 2. Photoionization cross-section of H$_2$@C$_{60}$ in HF and He@C$_{60}$ in HF and RPAE.

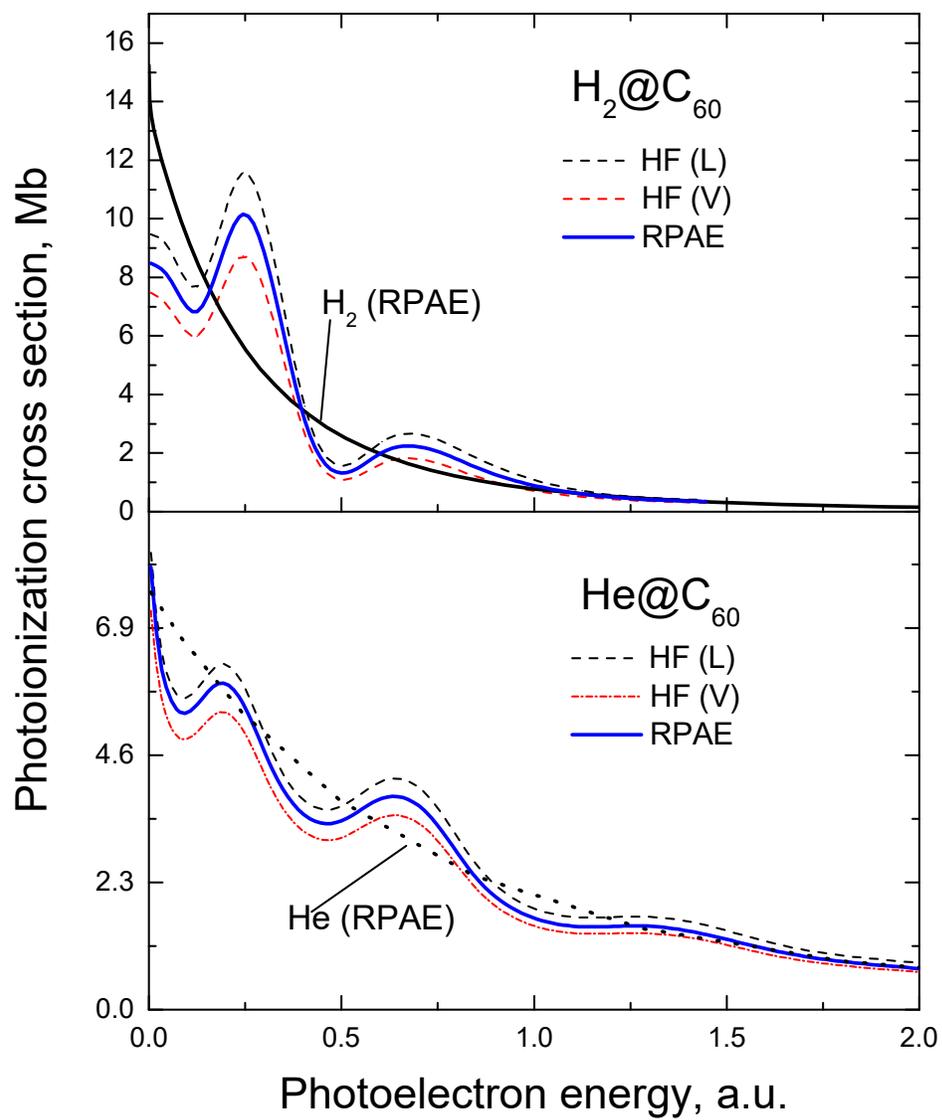

Fig. 3. Photoionization cross-section of H$_2$@C$_{60}$ and He@C$_{60}$ in HF and RPAE and H$_2$ and He

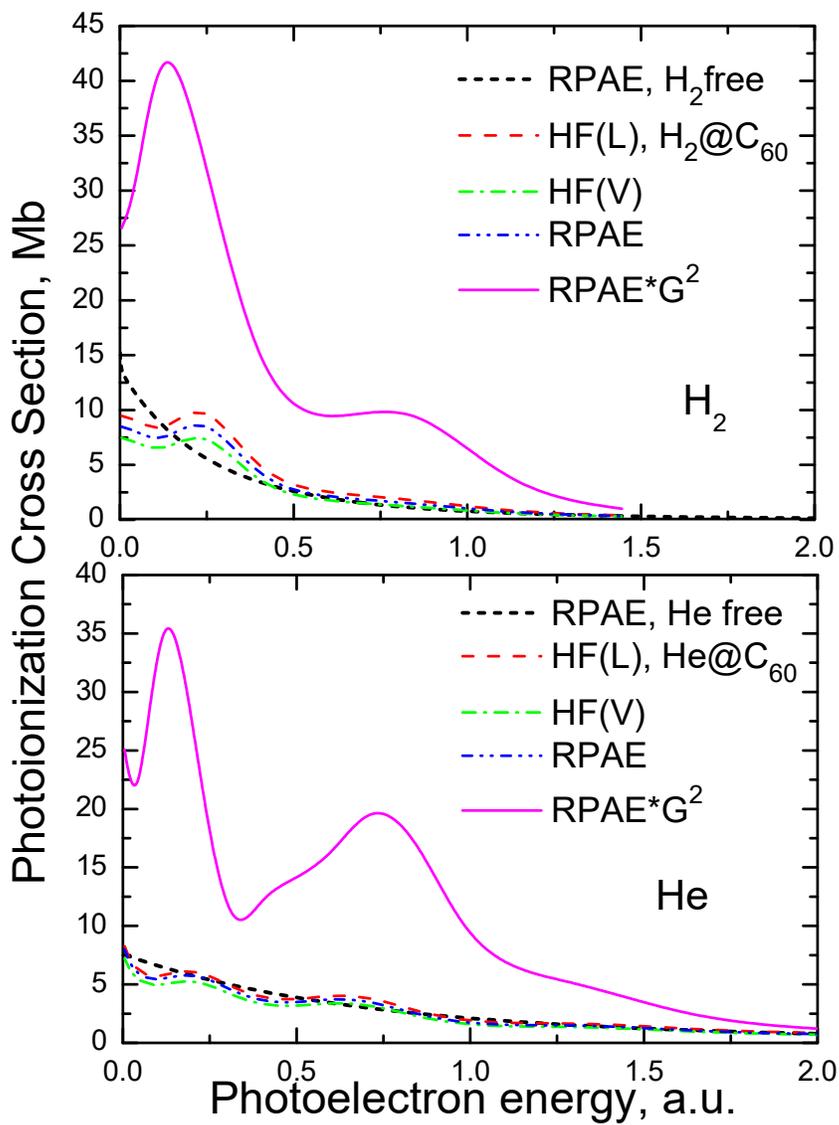

Fig. 4. Photoionization cross section of $H_2@C_{60}$ and $He@C_{60}$ in RPAE with account of fullerenes polarization